\documentclass[aps,twocolumn,superscriptaddress,amsmath,amssymb,showpacs]{revtex4}

\usepackage{graphicx}
\usepackage{indentfirst}
\usepackage{psfrag}
\usepackage{epsfig}
\usepackage{amsmath}
\usepackage{amssymb}
\usepackage{bm}
\usepackage{color}

\usepackage{booktabs}
\usepackage{longtable}
\usepackage{multirow}
\usepackage{ulem}

\def\be{\begin{eqnarray}}
\def\ee{\end{eqnarray}}

\newcommand{\Eg}{{\bf E}}

\newcommand{\pg}{{\bf p}}

\newcommand{\rg}{{\bf r}}

\newcommand{\ug}{{\bf u}}
\newcommand{\eg}{{\bf e}}

\newcommand{\Psig}{\mathbf{\Psi}}

\newcommand{\Gg}{{\bf G}}

\newcommand{\dd}{\mathrm{d}}

\begin{document}
\title{Speckle fluctuations resolve the interdistance between incoherent point sources \\ in complex media}

\author{R. Carminati}\email{remi.carminati@espci.fr}
\affiliation{ESPCI ParisTech, PSL Research University, CNRS, Institut Langevin, 1 rue Jussieu, F-75005, Paris, France}

\author{G. Cwilich}
\affiliation{Department of Physics, Yeshiva University, 500 W 185th Street, New York, New York 10033, USA}

\author{L.S. Froufe-P\'erez}
\affiliation{Department of Physics, University of Fribourg, Chemin du Mus\'{e}e 3, CH-1700, Fribourg, Switzerland}

\author{J.J. S\'aenz} \email{juanjo.saenz@uam.es}
\affiliation{Depto. de F\'{\i}sica de la Materia Condensada,  Instituto Nicol\'as Cabrera and Condensed Matter Physics Center (IFIMAC), Universidad
Aut\'{o}noma de Madrid, 28049 Madrid, Spain}
\affiliation{Donostia International Physics Center (DIPC), Paseo Manuel Lardizabal 4, 20018 Donostia-San Sebastian, Spain}

\begin{abstract}
We study the fluctuations of the light emitted by two identical incoherent point sources in a disordered environment. 
The intensity-intensity correlation function and the speckle contrast, obtained after proper temporal and configurational averaging, 
encode the relative distance between the two sources. 
This suggests the intriguing possibility that intensity measurements at only one point in a speckle pattern produced
by two incoherent sources can provide information about the relative distance between the sources, with a precision that is not limited
by diffraction. The theory also suggests an alternative approach to Green's function retrieval technique, where the correlations of the isotropic ambient noise 
detected by two receivers are replaced by a measurement at a single point of the noise due to two fluctuating incoherent  sources. 
\end{abstract}


\date{\today}

\maketitle

\section{Introduction}

Pushing the resolution limits of light microscopy, and understanding optical phenomena on scales below the diffraction limit, 
has been the driving force of what is known today as nano-optics~\cite{Novotny}. To overcome this limit, most of the early work was focused on near-field 
optical microscopy and related  techniques~\cite{Betzig}. However, in recent years, new concepts in fluorescence microscopy have  pushed the resolution 
of far-field imaging down to the nanometer range~\cite{Hell}. Most of these methods~\cite{varios} rely on the accurate localization of individual fluorescent 
markers, that are isolated from one another on the basis of one or more distinguishing optical characteristics, or by selective or random activation of a bright 
and a dark state~\cite{Hell}. Determining the location of an isolated fluorescent marker is only limited by photon noise, and not by the diffraction barrier.

A key issue affecting these subwavelength imaging methods is the optical transparency of the media surrounding the light emitters. Taking advantage 
of the transparency of cells, fluorescence microscopy uniquely provides noninvasive imaging of the interior of cells and allows the detection of specific 
cellular constituents through fluorescence tagging. However, certain biological tissues or soft-matter systems (such as foams or colloidal suspensions)
look turbid due to intense scattering of photons traveling through them~\cite{Yold}. The image formed at a given point in the observation plane consists in a 
superposition of multiple fields, each arising from a different scattering sequence in the medium. This gives rise to a chaotic intensity distribution
with numerous bright and dark spots known as a speckle pattern, producing a blurred image carrying no apparent information about 
the source position~\cite{speckle}.  

Techniques to measure the distance between individual nano-objects without actually imaging their position exist~\cite{Michalet},
Fluorescence Resonance Energy Transfer (FRET) being the most widespread example~\cite{FRET}. 
It relies on the near-field energy transfer between two fluorophores
(donor and acceptor) emitting at different wavelengths. The FRET signal ({\it e.g.} the ratio between the intensities emitted by the donor and the acceptor at different wavelengths) depends on the donor-acceptor distance in the range $2\sim10$~nm. As such, it is not very sensitive to scattering problems. 
However, determining distances between two emitters in the range of 10 to 500~nm in a scattering medium still remains a challenging problem, not accessible either 
by fluorescence microscopy or FRET techniques. 
Our main goal here is to introduce a new approach to obtain information about the relative distance between two identical incoherent point sources in a disordered environment, based on the analysis of the  fluctuations of the emitted light. 
This is an issue  of much interest, for example, in the study of conformational changes in biomolecules in living tissues. 
Sensing the distance between two incoherent sources in a complex medium could also provide an alternative to Green's function retrieval techniques based
on the correlations of the isotropic ambient noise measured at two receivers~\cite{GFR}.

In this paper, we propose a method to capture the interaction between two identical sources in a scattering environment, based only on the measurement
of intensity fluctuations. The principle of the method is schematically illustrated in Fig.~1, and is based on the analysis of the intensity-intensity correlation function and the intensity 
fluctuations in the speckle pattern formed by two identical and mutually incoherent point sources. 
This approach permits, in principle, to monitor the relative distance between the sources in the range 10-500 nm, with a precision
that is not limited by diffraction, but by the microstructure of the scattering medium. In application to Green's function retrieval in complex media, the approach
replaces the two-point field-field correlation of the background noise by a measurement at a single point of the intensity noise due to the two fluctuating sources. 
This might simplify the technique, in particular at visible or near-IR frequencies where time-domain field-field correlations are not easy to measure.
The result in this paper also illustrate the fact that multiple scattering, that had long been considered as an unavoidable nuisance, can actually enhance the performance of sensing, imaging and communication techniques~\cite{Fink}, as already demonstrated in the context of spatio-temporal focusing by time reversal~\cite{TR,Pierrat2007}, wavefront shaping 
of multiply scattered waves~\cite{wavefront}, or improvement of information capacity of telecommunication channels~\cite{telecom}.

\begin{figure}[htbp]
\begin{center}
\includegraphics[width=8cm]{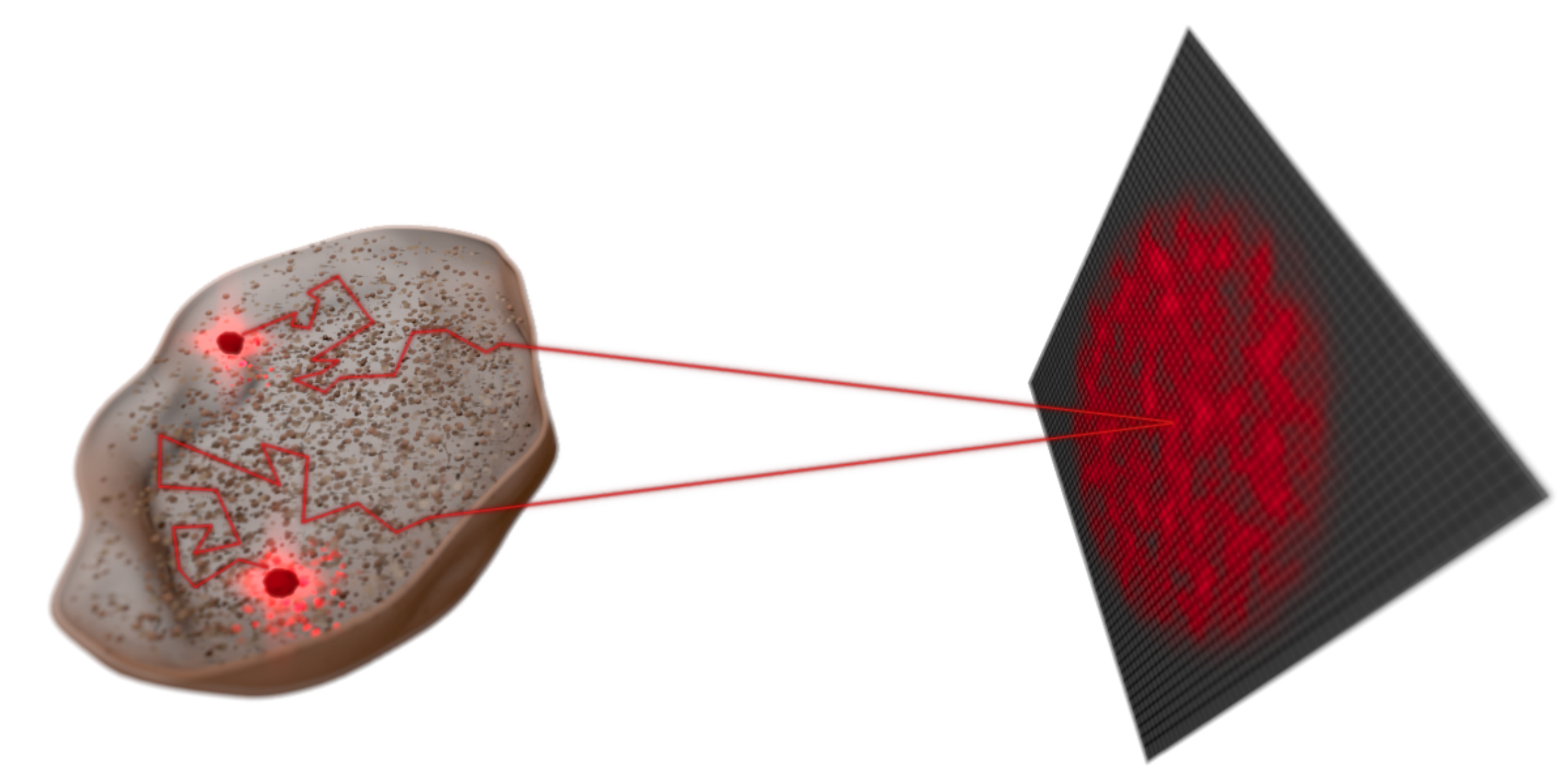}
\caption{The intensity radiated by two incoherent point sources in a complex medium form a speckle pattern that fluctuates
in both space and time. The speckle fluctuations encode the relative distance between the sources.}
\end{center}
\end{figure}

\section{Fluctuations in the power emitted by two incoherent sources}

We consider two point sources of light (electric dipoles) located at $\rg_1$ and $\rg_2$ in a disordered medium. The sources are characterized
by their electric dipole moments $\pg_1(t)$ and $\pg_2(t)$, that are fluctuating quantities of the form $\pg_k(t)=p_k \exp[i\phi_k(t)] \exp(-i\omega t) \ug_k$
with $\phi_k(t)$ a slowly varying random phase, $p_k$ a complex amplitude and $\ug_k$ a unit vector defining the orientation of the dipole moment.
This corresponds to a classical model for a quasi-monochromatic temporally incoherent source, such as a fluorescent source emitting
at frequency $\omega$. 
We assume that the two sources are uncorrelated (or mutually incoherent), so that $\overline{\exp[i\phi_1(t)]\exp[-i\phi_2(t)]}=0$, 
where the bar denotes averaging over the fluctuations of the sources. 
Using the (dyadic) Green function $\Gg(\rg,\rg^\prime,\omega)$ of the disordered medium, the
electric fields at any point $\rg$ can be written~:
\begin{equation}
\Eg(\rg) = \mu_0 \omega^2 \Gg(\rg,\rg_1,\omega) \pg_1 + \mu_0 \omega^2 \Gg(\rg,\rg_2,\omega) \pg_2 \ .
\label{eq:field_general}
\end{equation}
The intensity associated to this field is a time fluctuating and spatially varying quantity that forms a time-dependent speckle pattern.

Let us first consider the total power $P$ emitted by the two sources. It reads
\begin{equation}
P = \int_S \frac{\epsilon_0 \, c}{2} \, |\Eg(\rg)|^2 \, \dd S
\label{eq:def_P}
\end{equation}
where $S$ is a sphere with radius $R\to \infty$ that encloses the disordered medium and $c$ is the speed of light in vacuum.
For a {\it non absorbing} medium, the following relation can be derived from the vector form of Green's second identity~\cite{Morse_Book}
(the frequency dependence in the Green function is dropped for simplicity):
\begin{align}
\frac{\omega}{c}\int_S \Gg(\rg,\rg_1) \pg_1 \cdot \Gg^*(\rg,\rg_2) \pg_2^* \, \dd S = \pg_1 \cdot \mathrm{Im}[ \Gg(\rg_1,\rg_2)] \pg_2^* \ .
\label{eq:Green_identity}
\end{align}
From Eqs.~(\ref{eq:field_general}-\ref{eq:Green_identity}), we obtain
\begin{align}
P = \frac{\mu_0 \, \omega^3}{2} \sum_{j,j^\prime=1}^2 p_j \, p^*_{j^\prime} \, \mathrm{Im} G_{jj^\prime}
\label{eq:P2}
\end{align}
where the notation $\mathrm{Im} G_{jj^\prime} = \ug_j \cdot \mathrm{Im}[ \Gg(\rg_j,\rg_{j^\prime},\omega)] \ug_{j^\prime}$ has
been introduced for the sake of simplicity.

We first assume that a temporal averaging over the fluctuations of the sources can be performed, in one configuration of the disordered medium
(frozen disorder). The fluctuation time scale of the emitted power can be associated to the coherence time, as usually defined for partially coherent sources~\cite{MandelWolf}.
For fluorescent sources, this time is on the order of the lifetime $\tau$ of the excited state. For emission in the visible range, 
expected orders of magnitude are $\tau \sim 1-10$ ns for dye molecules or quantum dots, and $\tau \sim 1-100 \mu$s for rare-earth ions.
The averaged power $\overline{P}$ is simply the sum of the averaged power emitted by 
each source independently, since the terms with $j \neq j^\prime$ in Eq.~(\ref{eq:P2}) vanish upon time averaging. It reads as
\be
\overline{P}= \frac{\pi\omega^2}{4 \epsilon_0} \left ( |p_1|^2 \, \rho_{11}+ |p_2|^2 \, \rho_{22} \right )
\label{eq:Pmoy}
\ee
where $\rho_{jj}=[2\omega/(\pi c^2)] \mathrm{Im} G_{jj}$ is the electric part of the local density of states (LDOS) at point $\rg_j$~\cite{Joulain2003}.
However, a cross-term survives in the fluctuations of the total emitted power. Indeed, calculating the variance of $P$
from Eq.~(\ref{eq:P2}), one obtains~\cite{note_average}
\begin{equation}
\overline{P^2} - \overline{P}^2 =  
\frac{\mu_0^2 \, \omega^6}{4} \left [  2 |p_1|^2|p_2|^2 \, ( \mathrm{Im} G_{12})^2\right ] \ .
\label{eq:VarP_time}
\end{equation}

The imaginary part of the two-point Green function $\mathrm{Im} G_{12}$ in Eq.~\eqref{eq:VarP_time} is known to enter the expression
of field-field spatial correlations in random fields, such as blackbody radiation or volume speckle patterns~\cite{RytovBook,Joulain2005,Shapiro_1989,Vynck2014},
the description of time-reversed fields~\cite{Pierrat2007}, and is at the core of Green's function retrieval techniques based on ambient noise 
correlations~\cite{GFR}. It is proportional to the cross density of states (CDOS), that was introduced in a different context for the description
of spatial coherence in complex systems~\cite{Caze2013}. Physically, the CDOS counts the number of photonic eigenmodes connecting two
points (in our case the source points) at a given frequency~\cite{Sauvan2014}. 
More precisely, the CDOS connecting $\rg_k$ to $\rg_j$ is given by $\rho_{jk}=[2\omega/(\pi c^2)] \mathrm{Im} G_{jk}$~\cite{note1}.
Using the CDOS, and assuming that the two sources have the same amplitude ($p_1 = p_2 \equiv p$),
the variance of the total emitted power can be rewritten as
\begin{equation}
\overline{P^2} - \overline{P}^2  =  
\frac{\pi^2 \, \omega^4}{8 \, \epsilon_0^2} |p|^4  \rho_{12}^2 \ .
\label{eq:varP_time_2}
\end{equation}
This equation is the first result in this paper. It provides a direct relationship between 
 the temporal fluctuations of the total power emitted by two incoherent sources and the CDOS
connecting the source points in an arbitrary environment. This suggests that a retrieval of the amplitude of the CDOS (or equivalently of the imaginary 
of the Green function at two different points) is possible in a structured medium from a measurement of temporal fluctuations of the emitted power 
emerging from two incoherent sources. Such a measurement would resemble the Green function retrieval from ambient noise correlations, based on
the relationship between field-field correlations and the imaginary part of the Green function given by the fluctuation-dissipation theorem,
initially introduced in the context of electromagnetic thermal fluctuations~\cite{RytovBook,LandauBook}. The generality of this relationship,
also valid for field fluctuations in speckle patterns~\cite{Shapiro_1989}, has stimulated the development of Green's function retrieval techniques
in acoustics, seismology or with low-frequency electromagnetic waves~\cite{GFR}. In this approach, the statistical isotropic ambient noise is detected 
by two receivers, while in the method suggested here the Green function is encoded in the noise due to the fluctuations of the two sources.
The possibility to measure power fluctuations instead of field-field correlations might be an advantage for electromanetic Green's function retrieval in the visible
or near-IR frequency range. Equation~(\ref{eq:varP_time_2}) also shows that the power fluctuations encode the interdistance between the sources. At this stage,
since the CDOS $\rho_{12}$ is specific to the sample under study and unknown, changes in power fluctuations could reflect changes in the interdistance,
but the interdistance could not be determined without solving a difficult inverse problem. We will see how the problem can be simplified in the presence
of multiple scattering in a disordered medium by peforming an ensemble averaging over the configurations of disorder.

\section{Configurational averaging in a disordered medium}

We now assume that in addition to a temporal averaging over the fluctuations of the sources, an average over the configurations of the disordered
medium can be performed. A specific situation would be that of sources embedded in a dynamic medium, with configurational changes occurring
on a time scale much larger than the characteristic time of the fluctuations of the sources. 
An equivalent situation is that of sources moving inside a frozen disordered medium, also on a sufficiently large time scale, as schematically shown in Fig.~2(a).
If both the sources and the disordered medium are fixed  (i.e. the medium itself does not fluctuate), an artificial configurational averaging process could 
be induced by an external moving diffuser surrounding the medium, as shown in Fig.~2(b).
In all these situations both averaging processes can be performed independently and subsequently.
The (temporal) variance of the total emitted power averaged over the configurations of the disordered medium is readily obtained
from Eq.~\eqref{eq:varP_time_2}:
\begin{equation}
\left\langle \overline{P^2} \right\rangle - \left\langle \overline{P}^2  \right\rangle=  
\frac{\pi^2 \, \omega^4}{8 \, \epsilon_0^2} |p|^4 \langle  \rho_{12}^2 \rangle
\label{eq:varP_speckle}
\end{equation}
where we use the notation $\langle ... \rangle$ for configurational averaging. 

\begin{figure}[htbp]
\begin{center}
\includegraphics[width=8cm]{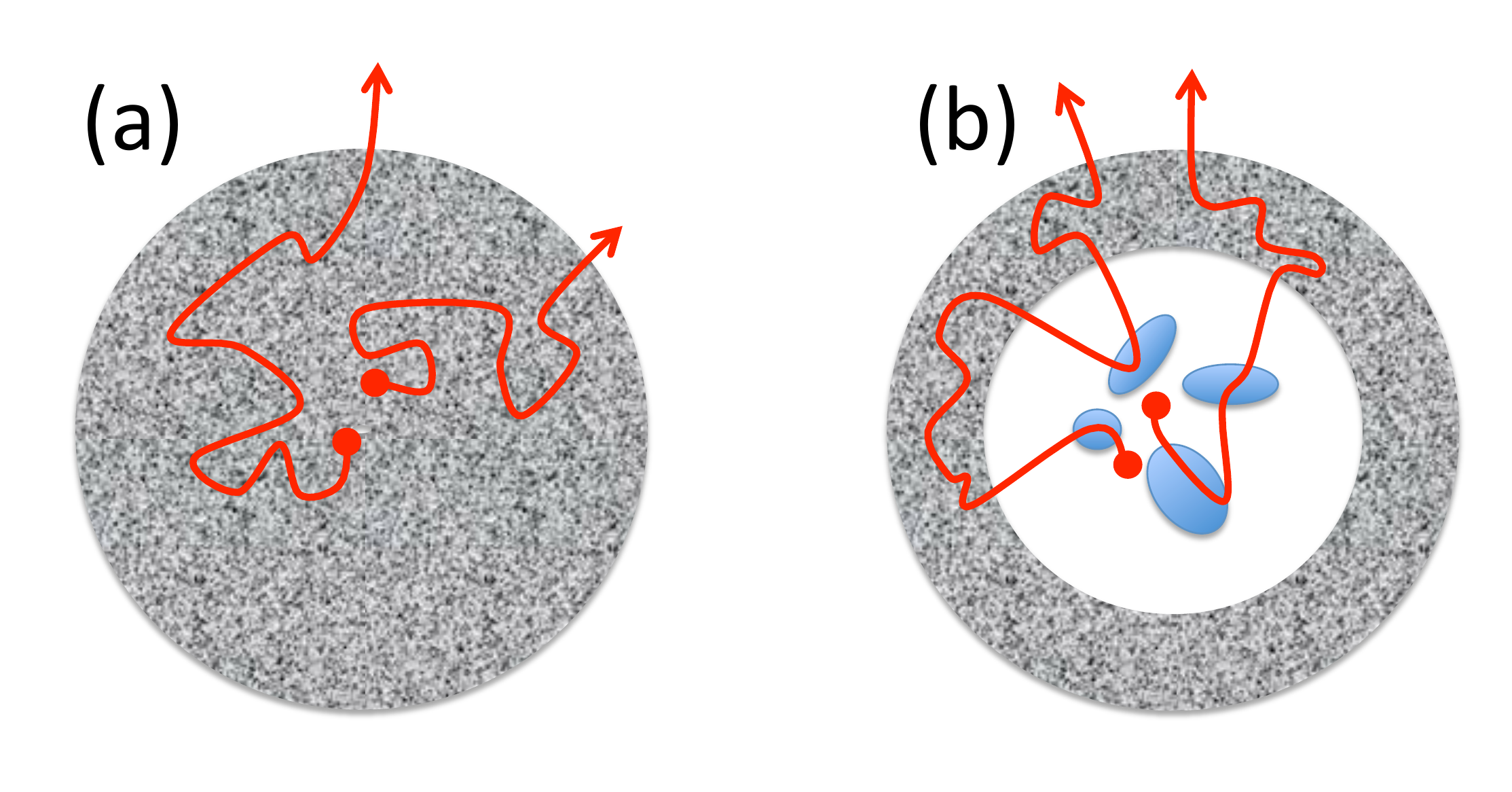}
\caption{Illustrations of two situations in which configurational averaging can be achieved. (a): Two sources embedded in a dynamic
scattering medium, or moving inside a static disordered medium. (b): The sources are pinned in a static medium and the speckle
is produced by an external diffuser surrounding the medium.}
\end{center}
\end{figure}

In practice, it is often convenient to work with normalized statistical quantities. 
For later convenience, we can introduce the contrast of the power fluctuations, that we define as
\be
\sigma_{P} \equiv \frac{\left\langle \overline{P^2} \right\rangle - \left\langle \overline{P}^2 \right\rangle}{\left\langle \overline{P} \right\rangle^2} \ .
\label{eq:def_sigmaP}
\ee
Let us point out again the specific use of two non-commuting averaging processes in this definition, one over the temporal fluctuations 
of the sources and subsequently one over the configurations of the disordered medium.
From Eqs.~\eqref{eq:Pmoy} and \eqref{eq:varP_speckle}, the power contrast can be written in terms of the LDOS and CDOS, leading to 
\be
\sigma_{P} =\frac{2\langle \rho_{12}^2 \rangle}{ ( \langle \rho_{11} \rangle + \langle \rho_{22} \rangle)^2} \ .
\label{eq:expr_sigmaP}
\ee
In the next section we will derive a simple relationship between the power contrast $\sigma_P$ and the speckle contrast $\sigma_S$ deduced
from the radiated intensity measured in a single direction, or equivalently at a single point in the far field speckle pattern.

\section{Speckle contrast}

A measurement of the total emitted power $P$ requires a detection integrated over $4\pi$ steradians.
Measuring  fluctuations of the intensity radiated in a given direction, or intensity correlations between two different  directions, 
could be a more convenient approach in practice. 
The time-averaged far-field intensity emerging in a given direction $\ug$, per unit solid angle, and for a polarization state $\eg_\alpha$, is given by
\be
\overline{I_{\alpha}(\ug)} &=& \overline{I_{1\alpha}(\ug) }+ \overline{I_{2\alpha}(\ug) } \label{inocros}
\ee
where $I_{j\alpha}(\ug)$ is the intensity radiated by the point source $\pg_j(t)$. The latter reads as
\be
I_{j\alpha}(\ug) = \lim_{r \to \infty} \frac{\mu_0 \omega^4}{2 c}\, r^2 \, |\eg_\alpha \cdot \Gg(\rg,\rg_j) p_j \ug_j|^2 \label{Ijsigma}
\ee
with $\rg = r \ug$. The configurational average of the (time averaged) total emitted power and fluctuations can be rewritten as 
\be
 \left\langle \overline{P} \right\rangle &=& \int_{4\pi}  \left\langle \overline{I(\ug) }  \right\rangle \, \dd\ug \label{exact1} \\
 \left\langle\overline{P^2}  \right\rangle  &=& \sum_{\alpha \alpha^\prime}
\iint_{4\pi}  \left\langle \overline{I_{\alpha}(\ug) I_{\alpha^\prime}(\ug^\prime)}  \right\rangle  \dd\ug \, \dd\ug^\prime 
\label{exact2}
\ee
where $\dd\ug$ means integration over the solid angle, $\alpha$ runs over the two orthogonal polarization states in the far field,
and $I(\ug) = \sum_{\alpha} I_\alpha(\ug)$.
We now assume that the radiated field is a random statistically isotropic field, under the only constraints given by 
Eqs.~\eqref{exact1} and \eqref{exact2}. In this case one has $\left\langle \overline{I(\ug) }\right\rangle  = \left\langle \overline{P}\right\rangle/(4\pi)$
and
\be
\left\langle \overline{I_{\alpha}(\ug) I_{\alpha^\prime}(\ug^\prime)}\right\rangle   =   \frac{1}{(8\pi)^2} \left\langle\overline{P^2}\right\rangle  
\left( 1 + \delta_{\alpha \alpha^\prime} \delta_{\ug \ug^\prime}\right)
\label{AvIp1Ip1}
\ee
where $\delta_{\alpha \alpha^\prime}$ is the usual Kronecker delta 
and $\delta_{\ug \ug^\prime}$ is a Kronecker delta with respect to detection angles. The derivation of Eq.~\eqref{AvIp1Ip1}
is given in Appendix A.

At  this stage, it is worth noticing that relation \eqref{AvIp1Ip1} between the speckle intensity-intensity correlation function and the
fluctuations of the total emitted power is the origin of  the so-called $C_0$ correlation known for a speckle
produced by a single source~\cite{Shapiro1999,Skipetrov2000,Skipetrov2006,Caze2010}.
If we consider the particular case of a single source at $\rg_1$ ($\pg_2=0$) that does not fluctuate in time, the normalized intensity-intensity
correlation function $C(\ug,\alpha;\ug^\prime,\alpha^\prime)$ follows directly from Eq. \eqref{AvIp1Ip1} and is given by
\be
C(\ug,\alpha;\ug^\prime,\alpha^\prime) &=& \frac{\langle I_{1\alpha}(\ug) I_{1\alpha^\prime}(\ug^\prime)\rangle}{\langle I_{1\alpha}(\ug) \rangle
\langle I_{1\alpha^\prime}(\ug^\prime) \rangle} -1 \nonumber \\ &=& C_0 + (C_0+1)\delta_{\alpha \alpha^\prime} \delta_{\ug \ug^\prime}
\label{C1}\ee
where
\be
C_0 &=& \frac{\langle \rho_{11}^2\rangle - \langle \rho_{11}\rangle^2}{\langle \rho_{11}\rangle^2}
\ee
is the normalized variance of the LDOS at the position of the source, that is at the origin of the infinite-range $C_0$ contribution
to the speckle correlation function~\cite{Skipetrov2006,Caze2010}. Therefore, for a single source, Eq.~\eqref{AvIp1Ip1} provides
another way of deriving the well-known results related to the $C_0$ correlation.
Moreover, from Eq. \eqref{AvIp1Ip1}, the speckle contrast (defined as the normalized variance of the intensity in a specific direction and a given polarization channel $\alpha$) is given by 
\be 
\frac{\left\langle I_{\alpha}(\ug)^2\right\rangle -  \left\langle I_{\alpha}(\ug)\right\rangle^2}{\left\langle I_{\alpha}(\ug)\right\rangle^2} = 1 + 2 C_0
\ee 
a result already obtained by Shapiro~\cite{Shapiro1999}, based on a microscopic diagrammatic approach for scalar waves. This result
leads to non-universal corrections to the Rayleigh statistical distribution of the intensity in speckle patterns produced by multiple scattering~\cite{Shapiro1999,Mirlin1998}.
 
For two fluctuating sources, assuming again that configurational changes in the disordered medium occur on time scales larger than the characteristic time of the 
source fluctuations, we also have [the derivation is similar to that leading to Eq.~(\ref{AvIp1Ip1}), see Appendix A]:
\begin{align}
& \left\langle \overline{I_{\alpha}(\ug)} \, \overline{I_{\alpha^\prime}(\ug^\prime)} \right\rangle = 
\frac{1}{(8\pi)^2}  \left\langle\overline{P}^2 \right\rangle + \frac{1}{(8\pi)^2}
\left\langle\overline{P^2}\right\rangle  \delta_{\alpha \alpha^\prime} \delta_{\ug \ug^\prime} \ .
\label{eq:fluct_IuP}
\end{align}
Making use of Eqs.~\eqref{eq:varP_speckle}, \eqref{AvIp1Ip1} and \eqref{eq:fluct_IuP} for two sources with the same amplitude $p$, we obtain
\begin{align}
\left\langle \overline{I_{\alpha}(\ug) I_{\alpha^\prime}(\ug^\prime)}  \right\rangle -  \left\langle \overline{I_{\alpha}(\ug)} \, \overline{I_{\alpha^\prime}(\ug^\prime)} \right\rangle
&= \frac{1}{(8\pi)^2} \left [ \left\langle\overline{P^2} \right\rangle - \left\langle\overline{P}^2 \right\rangle \right ] \nonumber \\
&= \frac{\omega^4}{8^3 \, \epsilon_0^2}|p|^4  \left\langle  \rho_{12}^2\right\rangle  \ .
\label{eq:fluct_Iu}
\end{align}
This equation shows that the intensities corresponding to two different speckle spots ($\ug \ne \ug^\prime$) in the angular speckle pattern 
formed by two incoherent point sources are strongly correlated, with a correlation given by the fluctuations $\langle  \rho_{12}^2\rangle$ of the CDOS.
We end up with the surprising result that for two incoherent sources a cross-term survives averaging, 
and induces infinite range correlations in the speckle pattern. 
This correlation is formally very similar to that given by the $C_0$ contribution in the case of a single source, the fluctuations
of the CDOS replacing the fluctuations of the LDOS. Since the CDOS is a two-point quantity, connecting in this specific situation
the two source points, the speckle correlations encode the distance between the two sources.

Another implication of Eq.~\eqref{eq:fluct_Iu} is that in the presence of multiple scattering, CDOS fluctuations
can be accessed from measurements of the directional intensity, and not only from measurements of the total power $P$
as suggested initially by Eq.~\eqref{eq:varP_speckle}.
More precisely, we will now show that the speckle contrast measured in a single speckle spot (hereafter denoted by
$\sigma_{S}$) contains the same information as the contrast $\sigma_{P}$ in Eq.~\eqref{eq:expr_sigmaP}, that assumed 
a measurement of the total emitted power. To proceed, we define the speckle contrast for a detection in a given polarization channel $\alpha$ as
\begin{align}
\sigma_S &\equiv \frac{\left\langle \overline{I_{{\alpha}}(\ug)^2}  \right\rangle -  \left\langle\overline{I_{{\alpha}}(\ug)}^2 \right\rangle}{\left\langle \overline{I_{{\alpha}}(\ug)}\right\rangle^2} 
\label{CTN2}
\end{align}
in a similar way as the contrast of the total emitted power in Eq.~\eqref{eq:def_sigmaP}.
From Eq.~\eqref{eq:fluct_Iu}, and the relation 
\be
\left\langle \overline{I_{{\alpha}}(\ug)}\right\rangle = \frac{\left\langle \overline{P} \right\rangle}{8\pi}
\ee
one immediatly obtains
\be 
\sigma_{S}= \frac{\left\langle \overline{P^2} \right\rangle - \left\langle \overline{P}^2 \right\rangle}{\left\langle \overline{P} \right\rangle^2} = \sigma_P
\label{SC2}
\ee 
or equivalently
\be
\sigma_{S} =  \frac{2\langle \rho_{12}^2 \rangle}{( \langle \rho_{11} \rangle + \langle \rho_{22} \rangle)^2} \ .
\label{SCfinal}
\ee
This equation is the second important result in this paper. It shows that for two incoherent sources in a disorder medium,
the speckle contrast measured in a single speckle spot, and computed from both temporal averaging over the fluctuations of the sources
{\it and} configurational averaging over disorder, is proportional to $\langle \rho_{12}^2 \rangle$ that characterizes the fluctuations of the CDOS 
connecting the two sources.This means that the information on the interdistance between the sources is encoded in the speckle contrast $\sigma_{S}$ 
measured at a single point. Changes in the speckle contrast could therefore be used to detect changes in the interdistance between the two sources. 
Compared to a measurement of the total emitted power $P$, a measurement
of the speckle contrast $\sigma_S$ would require a simpler instrumentation, but at the cost of a substantial reduction of the signal level. Note that a
parallel detection of the fluctuations in several speckle spots could be performed using a CCD camera. In practice, a compromise between simplicity
in instrumentation and signal-to-noise ratio should be found. 
The contrast in Eq.~(\ref{SCfinal}) would decrease to zero when increasing the interdistance. This change occurs 
on a range given by the width of the CDOS (Green function) considered as a function of the interdistance $r_{12}=|\rg_2-\rg_1|$. This width depends
on the microscopic structure of the disordered medium, providing in principle resolution capabilities beyond the free-space diffraction limit.
The determination of the absolute value of the interdistance from the speckle contrast would require an expression of $\langle \rho_{12}^2 \rangle$. 
The crucial issue in this case is to find the
expression of $\langle \rho_{12}^2 \rangle \sim \langle (\mathrm{Im}G_{12})^2 \rangle$ in a disordered medium. This can be done at least in the weak
scattering limit, as discussed in the next section. Another consequence of Eq.~(\ref{SCfinal}), or Eq.~\eqref{eq:fluct_Iu},
 is that it provides a way to measure $\langle (\mathrm{Im}G_{12})^2 \rangle$
in a complex medium from the speckle noise recorded at a single point, which in the context of Green's function retrieval might provide an alternative
to the two-point field-field correlation technique, as discussed previously in section II.

\section{Weak scattering limit}

When the sources are embedded in a weakly disordered, homogeneous and isotropic medium, it is possible to give an explicit approximate
expression of the fluctuations of the CDOS $\langle \rho_{12}^2 \rangle = [4\omega^2/(\pi^2 c^4)] \langle (\mathrm{Im} G_{12})^2 \rangle$. 
In the limit $k_0 n_{\text{eff}} \ell \gg 1$, where $n_{\text{eff}}$ is the effective refractive index, $k_0=2\pi/\lambda$ (with $\lambda$ the wavelength in vacuum) and $\ell$ the Boltzmann scattering mean free path,
the averaged Green function is given by the well known result for homogeneous and isotropic media and  
we can write 
$ \langle   \mathrm{Im} G_{jk}  \rangle $ as  
\begin{align}
\langle   \mathrm{Im} G_{jk}  \rangle \approx  \left( \ug_1.\ug_2 +\frac{(\ug_1.\bm{\nabla}_{\rg_1}) 
(\ug_2 .\bm{\nabla}_{\rg_2})}{k_{\text{eff}}^2} \right)  \frac{\sin\left(  k_{\text{eff}} \, r_{12}\right)}{4\pi r_{12}} \label{Ghom}
\end{align}
where $k_{\text{eff}} =  k_0 n_{\text{eff}} +i/(2\ell)$ is the effective wavenumber. 
The description of the scattering medium by a complex effective wavenumber breaks down when the distance $r_{12}$ between the emitters
approaches the size of the homogeneities (more precisely the correlation length of disorder, see the discussion in Ref. \cite{Carminati2009} and references therein). Although 
this might look like a severe limitation, this approximation is in practice very robust and has been shown to model accurately light diffusion through biological tissues.
To lowest order, we can further approximate 
$ \langle (\mathrm{Im} G_{jk})^2  \rangle \approx  \langle   \mathrm{Im} G_{jk}  \rangle^2$ which, from Eqs.~\eqref{eq:fluct_Iu}, or \eqref{SCfinal}, and \eqref{Ghom} provides an explicit expression of the intensity correlation function, or the speckle contrast, in terms of the relative distance between the sources. 
Such an explicit expression should be of practical interest, e.g., for the sensing of the interdistance between emitters embedded in biological tissues. 
Let us point out that in terms of the detection of {\it changes} in the interdistance (without measuring its absolute value), the method suggested in this paper
does not requite any explicit expression of $\langle \rho_{12}^2 \rangle$ and relies only on Eqs. (\ref{eq:varP_time_2}), (\ref{eq:expr_sigmaP}) or (\ref{SCfinal}).

\section{Conclusion}

The results derived in this paper suggest the intriguing possibility that intensity measurements at only one point in a speckle pattern produced
by two incoherent sources can provide information about the relative distance between the sources. Moreover, this information is in principle not limited
by diffraction. It can be extracted from the speckle contrast $\sigma_{S}$ obtained after a proper time and configurational averaging process.
The results also suggests an alternative approach to the Green function retrieval technique. In the later the statistical isotropic ambient noise is 
detected by two receivers, while in the method suggested here the Green function is encoded in the noise due to two fluctuating incoherent 
sources measured at a single point. Finally, let us note that for fluorescent emitters, the quantum optical equivalent of Eqs.~\eqref{eq:varP_time_2}, \eqref{eq:expr_sigmaP}
or \eqref {SCfinal} might be established in terms of decay rate or photocount statistics of a coupled system, and fluctuations
could be associated with super-radiant and sub-radiant states. This quantum treatment is left for future work. 

\acknowledgments

We acknowledge helpful discussions with  M. Fink and F. Scheffold. This work was supported by the Spanish MEC through Grant No.   FIS2012-36113, 
by the Comunidad de Madrid through Grant  No. S2009/TIC-1476 (Microseres Project), by LABEX WIFI 
(Laboratory of Excellence within the French Program "Investments for the Future") under references ANR-10-LABX-24 and ANR-10-IDEX-0001-02 PSL*.  JJS acknowledges  an IKERBASQUE Visiting Fellowship and 
RC and GC  the hospitality of the  DIPC at Donostia-San Sebastian (Spain) where part of this work was done.

\appendix

\section{Speckle intensity correlations: Derivation of Eq. \eqref{AvIp1Ip1} from  Random Matrix Theory}
\label{app}
The  far-field intensity radiated by two dipoles emerging in a given direction $\ug$, per unit solid angle, and for a polarization state $\eg_\alpha$, can be written as
\be
I_{\alpha}(\ug) = \lim_{r \to \infty} \frac{\mu_0 \omega^4}{2 c}\, r^2 \, \left| \sum_{j=1}^2 \eg_\alpha \cdot \Gg(\rg,\rg_j) \pg_j \right|^2 
\ee
with $\rg = r \ug$.
Assuming that the $4\pi$ solid angle is divided in a finite number  $N \gg 1$, of (solid angle) pixels, the total power  emitted   by the two sources is given by
\be
P &=& \int_{4\pi} I(\ug) \dd \, \ug =   \frac{4\pi}{N} \sum_\alpha \sum_u^N  I_\alpha(\ug) 
\ee
which can be written in compact matrix form,
\be
P &=& \frac{4\pi}{N} \hat{\pg}^\dagger \Psig^\dagger \Psig \hat{\pg} 
\ee  
where $\hat{\pg}^\dagger=(\pg^\dagger_1, \pg^\dagger_2) = (p^*_{1x},p^*_{1y},p^*_{1z},p^*_{2x},p^*_{2y},p^*_{2z})$ and $\Psig$ is a $2N \times 6$ matrix ($2N$ comes from the two orthogonal polarizations of the field for each observation angle in the far-field). By using the singular value decomposition (SVD), the $\Psig$ matrix can be factorized as
\be
\Psig& =& \mathbf{U}\mathbf{\Sigma}\mathbf{V}^\dagger \\
\Psig^\dagger \Psig &=& \mathbf{V} \mathbf{\Sigma}^2 \mathbf{V}^\dagger 
\ee
where $ \mathbf{U}$ is an unitary $2N \times 2N$ matrix,  $\mathbf{\Sigma}$ is a $2N \times 6$  rectangular diagonal matrix with nonnegative real numbers on the diagonal and   $ \mathbf{V}$ is a $6\times 6$ unitary matrix.  Equation \eqref{eq:P2} in the main text can now be written as 
\begin{align}
P &= \frac{4\pi}{N} [ \hat{\pg}^\dagger\Psig^\dagger \Psig \hat{\pg} ]= \frac{4\pi}{N} [ \hat{\pg}^\dagger\mathbf{V} \mathbf{\Sigma}^2 \mathbf{V}^\dagger \hat{\pg} ]  \nonumber \\ &=  \frac{\mu_0 \, \omega^3}{2} \sum_{j,j' =1}^2  p_{j'}^*p_j \mathrm{Im} G_{jj'}  \nonumber \\ &=  \frac{\mu_0 \, \omega^3}{2}  \begin{pmatrix} \pg_1^\dagger & \pg_2^\dagger \end{pmatrix} \begin{bmatrix}
\mathrm{Im}\Gg(11) & \mathrm{Im}\Gg(12) \\
\mathrm{Im}\Gg(21) & \mathrm{Im}\Gg(22)
\end{bmatrix} \begin{pmatrix} \pg_1 \\ \pg_2 \end{pmatrix}
\label{eq:mat1}
\end{align}
where $\Gg(jj') \equiv \Gg(\rg_j,\rg_{j'})$.
In terms of these new matrices, the intensity at a given angle $\ug$ and polarization $\alpha$ can be computed as
\begin{align}
I_\alpha(\ug) =  \sum_{a,b,a',b' =1}^{6}  \hat{p}_{a'}^\dagger V_{a' b'} \Sigma_{b'b'} U_{b' u_\alpha}^{ \dagger}U_{u_\alpha b} \Sigma_{bb}V_{ba}^\dagger  \hat{p}_a  \label{eq:Isum}
\end{align}
where $\hat{p}_{a}$ corresponds to the $i_a$ component of the dipole $\pg_{j_a}$ (with $j_a=\text{int}\Big[(a+2)/3\Big]$ and $i_a=a+3-3j_a$).

If we assume that the radiated field is statistically isotropic, we can consider $\mathbf{U}$ as a random unitary matrix statistically independent of the $\mathbf{\Sigma}$ and $\mathbf{V}$ matrices (isotropy hypothesis). This is one of the key assumptions in the macroscopic scaling approach of transport theory  \cite{Mello:1988,Mello:1990,Saenz:2003}. By using the averages over the unitary group 
(evaluated by Mello in ref. \cite{Mello:1990}), 
\be \langle  U_{b' u_\alpha}^{ \dagger}U_{u'_{\alpha'} b} \rangle = \frac{1}{2N} \delta_{b b'} \delta_{\ug \ug'} \delta_{\alpha \alpha'}\ee and the ensemble average of the intensity is given by
 \begin{align}
\langle I_\alpha(\ug) \rangle =  \frac{1}{2N} \sum_{a,b,a' =1}^6  \hat{p}_{a'}^\dagger \left\langle V_{a' b}  \Sigma_{bb}^2V_{ba}^\dagger \right\rangle  \hat{p}_a 
\end{align}
which, from Eq. \eqref{eq:mat1}, gives
\begin{align}
\langle I_\alpha(\ug) \rangle =  \frac{\mu_0 \, \omega^3}{16 \pi} \hat{\pg}^\dagger \begin{bmatrix}
\langle\mathrm{Im}\Gg \rangle  \end{bmatrix}\hat{\pg}=\frac{1}{2}\frac{\langle P \rangle}{4\pi}
\label{eq:mat2}
\end{align}
with $\hat{\pg} = \begin{pmatrix} \pg_1 \\ \pg_2 \end{pmatrix}$ and 
\begin{align}
\langle\mathrm{Im}\Gg \rangle = \begin{pmatrix}
\langle\mathrm{Im}\Gg(11)\rangle  & \langle\mathrm{Im}\Gg(12)\rangle \\
\langle\mathrm{Im}\Gg(21)\rangle & \langle\mathrm{Im}\Gg(22)\rangle
\end{pmatrix}
\end{align}
i.e., as expected, under the isotropy hypothesis, the intensity at a given angle is simply proportional to the average of the total radiated power, 
$\langle I(\ug) \rangle = 2 \langle I_\alpha(\ug) \rangle = \langle P\rangle /(4 \pi)$.
After averaging Eq.~(\ref{eq:Isum}) over the temporal fluctuations of
the sources we have $\overline{\hat{p}_{a}\hat{p}_{b}}= \delta_{j_a j_b}$   and $\overline{I_\alpha(\ug)}$ is simply the sum of the individual intensities [Eq. \eqref{inocros}] i.e. the non-diagonal boxes of $\langle\mathrm{Im}\Gg\rangle$ do not  contribute (no crosstalk term). The same applies for \begin{align} &\overline{\langle I_\alpha(\ug) \rangle} = \left\langle \overline{I_\alpha(\ug)} \right\rangle= \left\langle \overline{P}\right\rangle/(8\pi) = \overline{ \langle P\rangle }/(8\pi) \\
&=   \frac{\mu_0 \, \omega^3}{16 \pi} \begin{pmatrix} \pg_1^\dagger & \pg_2^\dagger \end{pmatrix} \begin{pmatrix}
\langle\mathrm{Im}\Gg(11) \rangle & \bf{0} \\
\bf{0} & \langle\mathrm{Im}\Gg(22) \rangle  \end{pmatrix} \begin{pmatrix} \pg_1 \\ \pg_2 \end{pmatrix}
\end{align} 

Let us now consider Eq. \eqref{exact2},
\be
\left\langle\overline{P^2}\right\rangle  &=&   \frac{(4\pi)^2}{N^2}\sum_{\alpha \alpha'}\sum_{\ug,\ug'}
\left\langle \overline{I_\alpha(\ug) I_{\alpha'}(\ug')}\right\rangle , \label{exact3}
\ee
with $ I_\alpha(\ug)$ given by Eq. \eqref{eq:Isum}. Computation of Eq. \eqref{exact3} involves again $\overline{\hat{p}_{a}\hat{p}_{b}}= \delta_{j_a j_b}$ together with averages of 4 elements of an unitary random matrix \cite{Mello:1990}
\begin{align}
 \langle  U_{b' u_\alpha}^{ \dagger}U_{u_\alpha b} U_{c' u'_{\alpha'}}^{ \dagger}U_{u'_{\alpha'} c} \rangle &= 
 \frac{ \delta_{bb'}\delta_{cc'} + \delta_{bc'} \delta_{cb'} \delta_{u_\alpha u'_{\alpha'} }}{4N^2-1} \nonumber \\ &-
  \frac{\delta_{bc'} \delta_{cb'} + \delta_{bb'}\delta_{cc'}  \delta_{u_\alpha u'_{\alpha'}}}{2N(4N^2-1)}.
\end{align}
It is easy now to find:
\begin{align}
\left\langle \overline{I_\alpha(\ug) I_{\alpha'}(\ug')}\right\rangle &=   \frac{N}{2(2N+1)}  \frac{1}{(4\pi)^2}\left\langle\overline{P^2}\right\rangle \Big( 1+\delta_{\ug \ug'}\delta_{\alpha \alpha'}\Big)
\label{eq:appendix1}
\end{align}
which, in the large $N$ limit, gives Eq.~\eqref{AvIp1Ip1} in the main text. 
Following a similar procedure, one also obtains
\begin{align}
\left\langle \overline{I_\alpha(\ug)} \, \overline{I_{\alpha'}(\ug')} \right\rangle &=   \frac{N^2}{4N^2-1}  \frac{1}{(4\pi)^2}\left\langle\overline{P}^2\right\rangle \Big( 1-\frac{\delta_{\ug \ug'}\delta_{\alpha \alpha'}}{2N}\Big) \nonumber \\
&+ \frac{N^2}{4N^2-1}  \frac{1}{(4\pi)^2}\left\langle\overline{P^2}\right\rangle \Big( \delta_{\ug \ug'}\delta_{\alpha \alpha'} -\frac{1}{2N}\Big)
\label{eq:appendix2}
\end{align}
which, in the large $N$ limit, gives Eq.~\eqref {eq:fluct_IuP}.


\begin{thebibliography}{99}
\bibitem{note_average}
In the calculation of $\overline{P^2}$, only terms for which the phase factors $\exp[i\phi_k(t)]$ cancel
out survive the time averaging process, since the two sources are assumed to be uncorrelated.
\end{thebibliography}
\end{document}